\documentclass[aps,prl,reprint]{revtex4-1}
\usepackage{graphicx}
\usepackage{sidecap}

\begin{document}


\title{Spin--flip Limited Exciton Dephasing in CdSe/ZnS Colloidal Quantum Dots}


\author{Francesco Masia, Nicol$\grave{\rm o}$ Accanto, Wolfgang Langbein, and Paola Borri}
\affiliation{Cardiff University School of Physics and Astronomy and
School of Biosciences, The Parade, Cardiff CF24 3AA, United Kingdom}


\date{\today}

\begin{abstract}
The dephasing time of the lowest bright exciton in CdSe/ZnS wurtzite
quantum dots is measured from 5\,K to 170\,K and compared with
density dynamics within the exciton fine structure using a sensitive
three-beam four-wave-mixing technique unaffected by spectral
diffusion. Pure dephasing via acoustic phonons dominates the initial
dynamics, followed by an exponential zero-phonon line dephasing of
109\,ps at 5\,K, much faster than the $\sim10$\,ns exciton radiative
lifetime. The zero-phonon line dephasing is explained by
phonon-assisted spin-flip from the lowest bright state to dark
exciton states. This is confirmed by the temperature dependence of
the exciton lifetime and by direct measurements of the bright-dark
exciton relaxation. Our results give an unambiguous evidence of the
physical origin of the exciton dephasing in these nanocrystals.
\end{abstract}



\maketitle

The optical properties of excitons in colloidal semiconductor
quantum dots (CQDs) of size comparable or smaller than the exciton
Bohr radius have been the subject of intensive research since many
years, also owing to the simplicity of colloidal synthesis and the
consequent easiness to practically engineer CQD size, shape,
composition and in turn quantum
confinement\,\cite{AlivisatosScience96}. One of the greatest hopes
with these nanostructures is to achieve atom-like
absorption/emission spectra and in turn ultralong exciton coherence,
attractive for applications ranging from cavity quantum
electrodynamics\,\cite{LeThomasNL06} to laser
technology\,\cite{KlimovScience2000}. Among the various material
types, CdSe CQDs were studied since early
days\,\cite{AlivisatosScience96}. Noticeably, their strong and
spectrally wide absorption combined with a narrow and size-tuneable
emission spectrum in the visible range has promoted their
application as biolabels\,\cite{MichaletScience05}, and in
photovoltaics, invigorating the interest in CQDs.

The excitonic level structure in CdSe CQDs has been treated
theoretically\,\cite{EfrosPRB96} and is non-trivial. Due to the
s-like conduction band, p-like valence band and the electron and
hole spin, the envelop ground state contains 12 sub-levels.
Spin-orbit coupling splits the p-like valence band according to the
hole total angular momentum $J$ into the $J=3/2$ and $J=1/2$ band,
the latter being at lower energies and thus not further considered.
The $J=3/2$ band itself is split by the hexagonal crystal structure
according to the angular momentum projection $M$ along the crystal
$c$-axis, into the heavy-hole $|M|=3/2$ and the light-hole $|M|=1/2$
(crystal-field splitting). The resulting levels are split further by
shape asymmetry and electron-hole exchange
interaction\,\cite{EfrosPRB96}. Assuming a shape which retains
cylindrical symmetry around $c$, excitons levels are classified by
the absolute value of the total angular momentum projection $|F|$
along $c$. For large sizes and small or oblate ellipticity, the
lowest exciton level is of heavy-hole character, two-fold degenerate
and dark with $|F|=2$ and lies few meV below a two-fold degenerate
bright level with $|F|=1$. Prolate (i.e. rod-shaped) shape asymmetry
counteracts the crystal-field splitting in CdSe, eventually pushing
the light-hole states below the heavy-hole states, and resulting in
a lower dark $F=0$ state, called $0^{\rm L}$, becoming the lowest
level below the bright $|F|=1$ level\,\cite{EfrosPRB96,ZhaoNL2007}.

The dephasing time and in turn the homogeneous linewidth of the
lowest bright ground-state exciton level in wurtzite CdSe CQDs was
first addressed experimentally by Schoenlein et
al.\,\cite{SchoenleinPRL93, MittlemanPRB94} in 2-4\,nm diameter
nanocrystals using transient three-beam four-wave mixing (FWM)
photon echo. Within the available dynamic range, an ultrafast
dephasing of $85-270$\,fs was found at low temperature and
attributed to acoustic phonon interactions. Subsequent spectral-hole
burning (SHB) experiments \cite{PalinginisPRB03} on larger CdSe/ZnS
core/shell CQDs with 9\,nm core diameter demonstrated a composite
homogeneous lineshape with a sharp zero-phonon line (ZPL) of
6\,${\mu}$eV linewidth full-width at half maximum (FWHM) at 2\,K,
corresponding to 200\,ps dephasing time, superimposed on a few meV
broad band due to phonon-assisted transitions. Such a composite
lineshape and a ZPL of $\sim10\,{\mu}$eV at low temperature were
confirmed by more recent studies on single CdSe/ZnS
CQDs\,\cite{FerneeJPCC08,BiadalaPRL09,CoolenPRL08}. On the other
hand, the radiative lifetime of the lowest bright ground-state
exciton is $\sim10$\,ns \cite{LabeauPRL03}, indicating that even a
200\,ps dephasing time is not radiatively limited. Single-dot PL as
well as SHB experiments are affected by spectral diffusion, a
variation of the CQD transition frequency over time from slow
fluctuations of the CQD environment. It was thus speculated in those
reports that the measured $\sim10{\mu}$eV linewidth was limited by
spectral diffusion and that a coherence lifetime of $\sim100$\,ps
was only a lower bound.

We previously demonstrated in self-assembled quantum dots that
transient resonant FWM overcomes not only the inhomogeneous
broadening in ensembles via the photon echo formation, but also
spectral diffusion\,\cite{BorriPRL01}. By combining directional
selection with heterodyne detection we achieved shot-noise limited
detection sensitivity and a large dynamic range
\,\cite{BorriJPCM07}. In this work we have applied this technique to
CdSe/ZnS CQD ensembles and, owing to its much greater dynamic range
compared to Ref.\,\cite{SchoenleinPRL93}, resolved the ZPL dephasing
of the lowest bright ground-state exciton level, not limited by
spectral diffusion.

\begin{figure}
\includegraphics*[width=8cm]{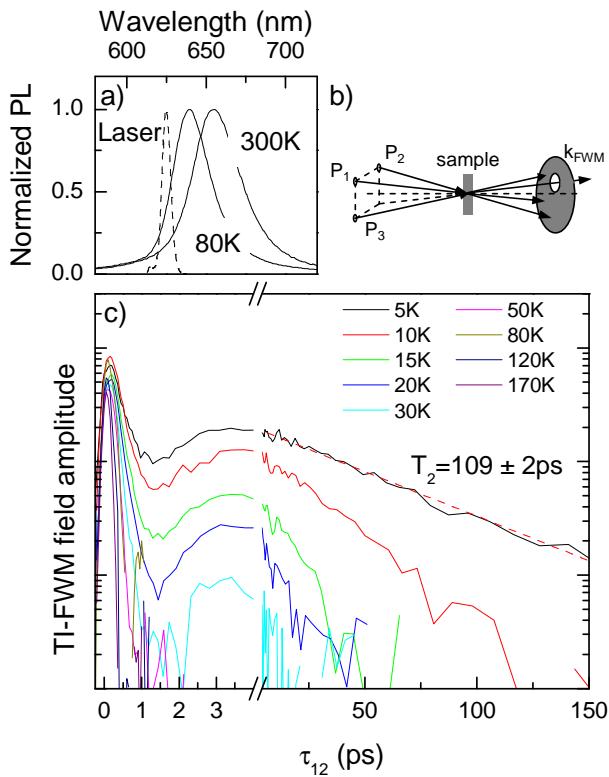}
\caption{(color online) a) Normalized PL spectra (solid lines) of
the investigated CdSe/ZnS wurtzite CQDs. The dashed line shows the
pulse laser spectrum used in the transient FWM experiment. b) Sketch
of the three-beam FWM directional geometry. c) Time--integrated FWM
field amplitude measured at different temperatures versus delay
between the first two pulses. The dashed line is a single
exponential fit to the data at 5\,K.\label{Fig1}}
\end{figure}

We investigated high-quality wurtzite CdSe/ZnS CQDs (Qdot655,
Invitrogen), nominally identical to those used in
Ref.\,\cite{FerneeJPCC08,BiadalaPRL09}, with a room temperature PL
peak emission wavelength at 655\,nm (see Fig.\,\ref{Fig1}a) and an
average core size of 8\,nm. High-resolution transmission electron
microscopy shows that the dots are typically non-spherical with a
rod-shaped core of $\sim5$\,nm diameter and $\sim10$\,nm length. The
CQDs were dispersed in a polystyrene film and sandwiched between two
quartz windows mounted onto a cold-finger cryostat. To
preferentially excite the lowest bright ground-state excitonic
absorption, the center wavelength of the exciting pulses was tuned
below the PL emission at low temperature by an amount of the order
of the Stokes shift ($\sim40$\,meV) in these CQDs. We used a
three-beam geometry, as sketched in Fig.\,\ref{Fig1}b, where each
beam is a train of 150\,fs pulses with 76\,MHz repetition rate. The
first pulse ($P_1$) induces a coherent polarization in the sample,
which after a delay $\tau_{12}$ is converted into a density grating
by the second pulse ($P_2$). The third pulse ($P_3$), delayed by
$\tau_{23}$ from $P_2$, is diffracted and frequency-shifted by this
density grating which is moving in the heterodyne
technique\,\cite{BorriJPCM07}, providing the FWM field which is
detected by its interference with a reference pulse of adjustable
delay. In an inhomogeneously broadened ensemble the FWM signal is a
photon echo emitted at $\tau_{12}$ after $P_3$ and the microscopic
dephasing is inferred from the decay of the photon echo amplitude
versus $\tau_{12}$. Conversely, the decay of the photon-echo
amplitude versus $\tau_{23}$ probes the exciton density
dynamics\,\cite{ShahBook96}. We recently applied this technique to
compare the dephasing time and exciton density dynamics in PbS
CQDs\,\cite{MasiaPRB11}. To minimize selective excitation of
linearly polarised transitions in the ensemble of randomly oriented
CQDs, all pulses were co-circularly polarised.

The time-integrated FWM (TI-FWM) field amplitude is shown in
Fig.\,\ref{Fig1}c as a function of $\tau_{12}$ for different
temperatures. Measurements are taken for non-zero $\tau_{23}=1$\,ps
to avoid non-resonant nonlinearities. The time-averaged excitation
intensity of $\sim$60\,W/cm$^{2}$ per beam was well within the
third-order nonlinear regime, and also resulted in negligible local
heating as we affirmed by power-dependent measurements. Remarkably,
the dephasing is initially very fast even at 5\,K, but within the
dynamic range of 3 orders FWM field corresponding to {\it 6 orders
FWM intensity} a long mono-exponential dephasing time ($T_2$) is
resolved at larger $\tau_{12}$. With increasing temperature the
initial dephasing becomes faster and more dominant, while
simultaneously $T_2$ decreases. This behavior reflects in time
domain the composite homogeneous lineshape consisting of a sharp
Lorentzian ZPL (corresponding to the long exponential dephasing)
superimposed onto a broad acoustic phonon band (the initial fast
dephasing), also observed in self-assembled quantum
dots\,\cite{BorriPRB05}. It is due to the excitation of localized
carriers, which distorts the lattice equilibrium and thus couples
the optical transition with phonon absorption/emission processes
similar to roto-vibrational bands in molecules. With increasing
temperature, it is expected that the weight of the ZPL (defined as
the area fraction of the ZPL in the linear absorption spectrum)
decreases and the width of the phonon band
increases\,\cite{BorriPRB05}.

\begin{figure}
\includegraphics*[width=8.5cm]{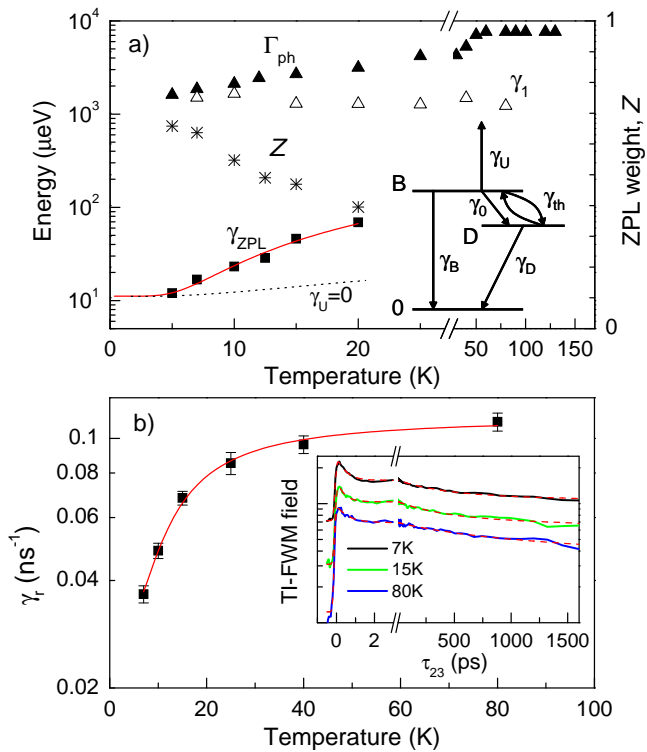}
\caption{a) Homogenous linewidth FWHM of the ZPL ($\gamma_{\rm
ZPL}$, squares) and of the acoustic phonon band ($\Gamma_{\rm ph}$,
triangles) versus temperature in CdSe/ZnS wurtzite CQDs. The ZPL
weight $Z$ is also shown (stars). The inset is a sketch of the lower
bright-dark exciton relaxation model. The solid line onto
$\gamma_{\rm ZPL}$ is a fit to the data (see text). The dotted line
is a fit without the absorption rate.  b) Long-lived exciton
lifetime versus temperature as measured from the exciton density
dynamics (see inset) by TI-FWM versus $\tau_{23}$ at $\tau_{12}=0$.
The solid line is a fit to the data. Dashed lines in the inset are
multi-exponential fits to the dynamics. $\gamma_{1}$ in (a) is the
broadening estimated from the initial subpicosecond density dynamics
(see text). \label{Fig2}}
\end{figure}

From the dynamics in Fig.\,\ref{Fig1}, we have quantified in
Fig.\,\ref{Fig2}a the temperature dependence of the ZPL width
$\gamma_{\rm ZPL}=2\hbar/T_2$, the FWHM of the acoustic phonon band
$\Gamma_{\rm ph}$, and the ZPL weight $Z$, using the method
discussed in Ref.\,\cite{BorriPRB05}. $\Gamma_{\rm ph}$ is obtained
by an exponential fit to the initial FWM decay which is an
approximation since the phonon broadband is non-Lorentzian but is in
qualitative agreement with what obtained by Fourier transforming the
FWM dynamics\,\cite{BorriPRL01}. Above 80\,K the initial dephasing
becomes faster than the pulse duration in the experiment hence the
reported $\Gamma_{\rm ph}$ represents a lower bound. Besides the
expected behavior of $Z$ and the acoustic phonon band, we observe a
decrease of $\gamma_{\rm ZPL}$ with decreasing temperature.
Importantly, the measured $T_2=109\pm2$\,ps ($\gamma_{\rm
ZPL}=12$\,$\mu$eV) at 5\,K not limited by spectral diffusion clearly
shows that the ZPL dephasing is far from the $\sim10$\,ns radiative
limit, which raises the question of its physical origin.

It was shown in Ref.\,\cite{LabeauPRL03} by temperature dependent
time-resolved PL in single wurtzite CdSe/ZnS CQDs that rapid
thermalization occurs between the lowest bright and the lowest dark
levels, following a three-level model as sketched in the inset of
Fig.\,\ref{Fig2}a. Within this model, after photoexcitation into the
bright state (B) rapid phonon-assisted relaxation occurs into the
dark state (D) with a low-temperature spin-flip rate
$\gamma_0\sim10$\,ns$^{-1}$ which is much larger than the radiative
recombination rate of the bright state $\gamma_{\rm
B}\sim0.1$\,ns$^{-1}$. With increasing temperature, not only
spontaneous emission but also stimulated emission and absorption of
acoustic phonons with energy equal to the bright-dark splitting
${\Delta}E_{\rm D}$ occurs with a rate $\gamma_{\rm
th}=\gamma_{0}N_{\rm B}$ where $N_{\rm B}=1/[\exp({\Delta}E_{\rm
D}/k_{\rm B}T)-1]$ is the phonon occupation number. For $k_{\rm
B}T\ll{\Delta}E_{\rm D}$ radiative recombination takes place mainly
from the dark state with a long lifetime $(\gamma_{\rm
D})^{-1}\sim1$\,${\mu}$s\,\cite{LabeauPRL03}. For $k_{\rm
B}T\gg{\Delta}E_{\rm D}$ thermalization between bright and dark
states results in a higher occupation probability of the bright
state. Exciton recombination therefore occurs from both levels and
the total decay rate is the thermal average between the two rates.
Since $\gamma_{0}\gg\gamma_{\rm B}$, such a spin-flip could be the
origin of the non-radiatively limited $\gamma_{\rm ZPL}$.

To prove this mechanism in our sample, we have first investigated
the temperature dependence of the exciton decay rate by measuring
FWM versus $\tau_{23}$ (see Fig.\,\ref{Fig2}b). The probed exciton
density exhibits a multi-exponential decay which we fit with three
time constants in the sub-picosecond, hundreds of picoseconds and
few nanosecond range (see inset of Fig.\,\ref{Fig2}b). The first
subpicosecond time constant is attributed to a sub-ensemble of CQDs
resonantly excited in the upper bright states, showing a rapid
relaxation towards the lower states. This time constant is not
temperature dependent up to 80\,K, consistently with $\sim$20\,meV
crystal field splitting\,\cite{EfrosPRB96} and correspondingly
negligible phonon occupation up to 80\,K. Such finding however
implies that the initial dephasing in Fig.\,\ref{Fig1}c might
include this relaxation. To quantify this effect we have plotted in
Fig.\,\ref{Fig2}a the homogeneous broadening $\gamma_1$ resulting
from the sub-ps relaxation from the upper-bright states. We find
that $\Gamma_{\rm ph}>\gamma_1$ above 7\,K indicating that the
initial dephasing is dominated by the acoustic phonon broadband at
these temperature. However at 5\,K due to this contribution we could
be underestimating the ZPL weight by about 10\% (as deduced from the
$\sim30$\% amplitude drop of the sub-ps density dynamics and the
cubic relationship between FWM field and $Z$\,\cite{BorriPRB05}).
The second time constant of $\sim400$\,ps found in the density
dynamics is temperature independent and attributed to Auger
recombination of charged excitons. The longest time constant,
manifesting as pulse to pulse pile-up included in our fit,
corresponds to a recombination rate $\gamma_{\rm r}$ which increases
with increasing temperature (see Fig\,\ref{Fig2}b), consistent with
Ref.\,\cite{LabeauPRL03}.

The three-level model sketched in the inset of Fig.\,\ref{Fig2}a
explains the temperature dependence of both $\gamma_{\rm ZPL}$ and
$\gamma_{\rm r}$. In this model, the dephasing rate is given by
$\gamma_{\rm ZPL}=\gamma_{\rm B} + \gamma_0 + \gamma_{\rm th}+
\gamma_{\rm U}$ where $\gamma_{\rm B}$ is the radiatively limited
dephasing of the bright state, $\gamma_0 + \gamma_{\rm th}$ is the
spin-flip relaxation into the dark state via spontaneous and
stimulated phonon emission, and $\gamma_{\rm U}$ accounts for phonon
absorption into an upper level. $\gamma_{\rm r}$ is the thermal
average of the radiative rates of the bright, the lower dark and the
upper level with their occupation probabilities and degeneracy. By
performing a combined fit of the measured temperature dependencies
of $\gamma_{\rm ZPL}$ and $\gamma_{\rm r}$ we find that the
dephasing rate is dominated by $\gamma_0 + \gamma_{\rm U}$ while the
stimulated emission term $\gamma_{\rm th}$ is actually negligible up
to 20\,K (see dotted line in Fig.\,\ref{Fig2}) considering the
bright-dark splitting ${\Delta}E_{\rm D}=2.0 \pm 0.2$\,meV found
from the fit of $\gamma_{\rm r}$ (see below). From the fit of
$\gamma_{\rm ZPL}$ we deduce $\gamma_0 = 11 \pm 0.5\, \mu
\textrm{eV}$ (corresponding to 17\,ns$^{-1}$ spontaneous spin-flip
rate) and a phonon absorption $\gamma_{\rm U}$ with amplitude
$130\pm40$\,$\mu$eV and activation energy $\Delta E_{\rm U}=2.2 \pm
0.3$\,meV. In prolate CQDs theory predicts \,\cite{EfrosPRB96} that
the dark level $0^{\rm L}$ approaches the lowest bright $|F|=1$,
with a cross over close to the $\sim5$\,nm diameter and $\sim10$\,nm
length \,\cite{ZhaoNL2007} mostly found in our sample. We thus
tentatively attribute this upper level to the $0^{\rm L}$ dark
state\,\cite{NoteCdSe}. The measured temperature dependence of
$\gamma_{\rm r}$ is well described by $\gamma_{\rm r}=(2\gamma_{\rm
B}+2\gamma_{\rm D}\exp({\Delta}E_{\rm D}/k_{\rm B}T)+\gamma_{\rm
D}\exp(-{\Delta}E_{\rm U}/k_{\rm B}T))/(2+2\exp({\Delta}E_{\rm
D}/k_{\rm B}T)+\exp(-{\Delta}E_{\rm U}/k_{\rm B}T))$ where for
simplicity we have assumed the radiative rate of the dark state
$\gamma_{\rm D}$ to be equal for the lower and upper dark levels. We
find that the temperature dependence is dominated by the
thermalization into the lowest dark state, hence the fit to the
measured $\gamma_{\rm r}$ is mostly sensitive to the lowest
bright-dark splitting ${\Delta}E_{\rm D}=2.0 \pm 0.2$\,meV with
$\gamma_{\rm B} = 2.4 \pm 0.1 \times 10^{-1}
\,\textrm{ns}^{-1}$\,\cite{NoteCdSe2}. Importantly, $\gamma_{\rm
B}\ll\gamma_{0}$, thus the zero-temperature extrapolated
$\gamma_{\rm ZPL}$ is indeed $\gamma_{0}$ limited by the spin-flip
rate.

\begin{figure}
\includegraphics*[width=8cm]{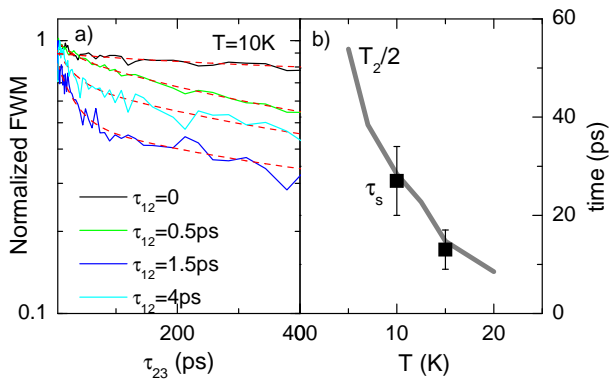}
\caption{a) TI-FWM field versus $\tau_{23}$ for different
$\tau_{12}$ together with fits (dashed lines) to the dynamics. b)
The additional time constant $\tau_{\rm s}$ (squares) inferred from
the dynamics at $\tau_{12}\neq0$ is compared with the measured ZPL
dephasing time. \label{Fig3}}
\end{figure}

To further verify this attribution we directly determined the
bright-dark relaxation by measuring for $\tau_{12}>0$ the FWM
amplitude versus $\tau_{23}$, which is sensitive to the density
dynamics within the fine-structure since a spectrally modulated
density grating is created. Specifically for the dark-bright
relaxation, one can understand this as follows. At $\tau_{12}=0$ the
creation of an exciton by the $P_1$ and $P_2$ pulses makes the
sample transparent to the probe $P_3$, since stimulated emission at
the ground-state to bright exciton transition (0-B) and absorption
at the exciton to biexciton transition (B-XX) compensate (note that
the biexciton binding energy $E_{\rm XX}$ is smaller that the
spectral width of the pulses). If the exciton is in the dark state,
the sample is also transparent to $P_3$ since both absorption and
stimulated emission is forbidden. We are thus not sensitive to the
dark-bright relaxation for $\tau_{12}=0$. Conversely, for
$\tau_{12}>0$ the phase difference between the fields from the 0-B
and B-XX transitions at the photon-echo time is $\pi+ E_{\rm
XX}\tau_{12}/\hbar$, resulting in exciton-biexciton beats versus
$\tau_{12}$\,\cite{ShahBook96,LangbeinPRB97}. We indeed observe weak
oscillations in the TI-FWM versus $\tau_{12}$ with a modulation
period in the 3-6\,ps range. In particular for $\tau_{12}$ equal to
half the exciton-biexciton beat period, 0-B and B-XX fields
interfere constructively, as opposed to the destructive interference
at $\tau_{12}=0$. A spin-flip into a dark level instead reduces the
field to zero. A bright-dark relaxation thus manifests as a decrease
in the FWM signal versus $\tau_{23}$, most pronounced for
$\tau_{12}$ equal to half the exciton-biexciton beat period. In
Fig.\,\ref{Fig3} the measured FWM amplitude versus $\tau_{23}$ at
10\,K is shown for three different values of $\tau_{12}$, and the
dashed lines are fits using a bi-exponential decay, plus an offset
accounting for the long radiative lifetime. The bi-exponential decay
consists of a new time-constant ($\tau_{\rm s}$) present only for
$\tau_{12}>0$, and the 400\,ps time constant previously discussed.
We see that the effect of $\tau_{\rm s}$ is most pronounced for
$\tau_{12}=1.5$\,ps and we can consistently fit all dynamics for
$\tau_{12}\neq0$ with $\tau_{\rm s}=27 \pm 7$\,ps at 10\,K. This
spin-flip time is in very good agreement with the density lifetime
$T_{2}/2$ deduced from the measured ZPL dephasing, as shown in
Fig.\,\ref{Fig3}b. Similar measurements carried out at 15\,K confirm
this agreement.

In conclusion, we have shown that the intrinsic zero-phonon line
dephasing of the ground-state exciton in CdSe/ZnS wurtzite colloidal
quantum dots is limited even at low temperatures by the rapid
($\sim100$\,ps) spin-flip from the lower bright to dark exciton
levels. This finding is different from InGaAs self-assembled quantum
dots, where the bright-dark exciton relaxation is longer than the
radiative lifetime and the exciton dephasing is radiatively
limited\,\cite{LangbeinPRB04a}. Our results conclusively resolve the
long standing question of the physical origin of the ZPL broadening
in CdSe/ZnS quantum dots.

\begin{acknowledgments}
We acknowledge discussions on the excitonic fine structure with Iwan
Moreels at Ghent University (Belgium). F.M. acknowledges the
European Union (Marie Curie grant agreement PIEF-GA-2008-220901) and
the Welcome Trust (VIP award). P.B. is a Leadership fellow of the
EPSRC UK Research Council (grant n. EP/I005072/1).
\end{acknowledgments}


\begin{thebibliography}{22}%
\makeatletter
\providecommand \@ifxundefined [1]{%
 \@ifx{#1\undefined}
}%
\providecommand \@ifnum [1]{%
 \ifnum #1\expandafter \@firstoftwo
 \else \expandafter \@secondoftwo
 \fi
}%
\providecommand \@ifx [1]{%
 \ifx #1\expandafter \@firstoftwo
 \else \expandafter \@secondoftwo
 \fi
}%
\providecommand \natexlab [1]{#1}%
\providecommand \enquote  [1]{``#1''}%
\providecommand \bibnamefont  [1]{#1}%
\providecommand \bibfnamefont [1]{#1}%
\providecommand \citenamefont [1]{#1}%
\providecommand \href@noop [0]{\@secondoftwo}%
\providecommand \href [0]{\begingroup \@sanitize@url \@href}%
\providecommand \@href[1]{\@@startlink{#1}\@@href}%
\providecommand \@@href[1]{\endgroup#1\@@endlink}%
\providecommand \@sanitize@url [0]{\catcode `\\12\catcode
`\$12\catcode
  `\&12\catcode `\#12\catcode `\^12\catcode `\_12\catcode `\%12\relax}%
\providecommand \@@startlink[1]{}%
\providecommand \@@endlink[0]{}%
\providecommand \url  [0]{\begingroup\@sanitize@url \@url }%
\providecommand \@url [1]{\endgroup\@href {#1}{\urlprefix }}%
\providecommand \urlprefix  [0]{URL }%
\providecommand \Eprint [0]{\href }%
\@ifxundefined \urlstyle {%
  \providecommand \doi  [0]{\begingroup \@sanitize@url \@doi}%
  \providecommand \@doi [1]{\endgroup \@@startlink {\doibase
  #1}doi:\discretionary {}{}{}#1\@@endlink }%
}{%
  \providecommand \doi  [0]{doi:\discretionary{}{}{}\begingroup
  \urlstyle{rm}\Url }%
}%
\providecommand \doibase [0]{http://dx.doi.org/}%
\providecommand \Doi [0]{\begingroup \@sanitize@url \@Doi }%
\providecommand \@Doi  [1]{\endgroup\@@startlink{\doibase#1}\@@Doi}%
\providecommand \@@Doi [1]{#1\@@endlink}%
\providecommand \selectlanguage [0]{\@gobble}%
\providecommand \bibinfo  [0]{\@secondoftwo}%
\providecommand \bibfield  [0]{\@secondoftwo}%
\providecommand \translation [1]{[#1]}%
\providecommand \BibitemOpen [0]{}%
\providecommand \bibitemStop [0]{}%
\providecommand \bibitemNoStop [0]{.\EOS\space}%
\providecommand \EOS [0]{\spacefactor3000\relax}%
\providecommand \BibitemShut  [1]{\csname bibitem#1\endcsname}%
\bibitem [{\citenamefont {Alivisatos}(1996)}]{AlivisatosScience96}%
  \BibitemOpen
  \bibfield  {author} {\bibinfo {author} {\bibfnamefont {A.}~\bibnamefont
  {Alivisatos}},\ }\href@noop {} {\bibfield  {journal} {\bibinfo  {journal}
  {Science},\ }\textbf {\bibinfo {volume} {271}},\ \bibinfo {pages} {933}
  (\bibinfo {year} {1996})}\BibitemShut {NoStop}%
\bibitem [{\citenamefont {Thomas}\ \emph {et~al.}(2006)\citenamefont {Thomas},
  \citenamefont {Woggon}, \citenamefont {Sch{\"o}ps}, \citenamefont {Artemyev},
  \citenamefont {Kazes},\ and\ \citenamefont {Banin}}]{LeThomasNL06}%
  \BibitemOpen
  \bibfield  {author} {\bibinfo {author} {\bibfnamefont {N.~L.}\ \bibnamefont
  {Thomas}}, \bibinfo {author} {\bibfnamefont {U.}~\bibnamefont {Woggon}},
  \bibinfo {author} {\bibfnamefont {O.}~\bibnamefont {Sch{\"o}ps}}, \bibinfo
  {author} {\bibfnamefont {M.~V.}\ \bibnamefont {Artemyev}}, \bibinfo {author}
  {\bibfnamefont {M.}~\bibnamefont {Kazes}}, \ and\ \bibinfo {author}
  {\bibfnamefont {U.}~\bibnamefont {Banin}},\ }\href@noop {} {\bibfield
  {journal} {\bibinfo  {journal} {Nano Lett.},\ }\textbf {\bibinfo {volume}
  {6}},\ \bibinfo {pages} {557–561} (\bibinfo {year} {2006})}\BibitemShut
  {NoStop}%
\bibitem [{\citenamefont {Klimov}\ \emph {et~al.}(2000)\citenamefont {Klimov},
  \citenamefont {Mikhailovsky}, \citenamefont {Xu}, \citenamefont {Malko},
  \citenamefont {Hollingsworth}, \citenamefont {Leatherdale}, \citenamefont
  {Eisler},\ and\ \citenamefont {Bawendi}}]{KlimovScience2000}%
  \BibitemOpen
  \bibfield  {author} {\bibinfo {author} {\bibfnamefont {V.~I.}\ \bibnamefont
  {Klimov}}, \bibinfo {author} {\bibfnamefont {A.~A.}\ \bibnamefont
  {Mikhailovsky}}, \bibinfo {author} {\bibfnamefont {S.}~\bibnamefont {Xu}},
  \bibinfo {author} {\bibfnamefont {A.}~\bibnamefont {Malko}}, \bibinfo
  {author} {\bibfnamefont {J.~A.}\ \bibnamefont {Hollingsworth}}, \bibinfo
  {author} {\bibfnamefont {C.~A.}\ \bibnamefont {Leatherdale}}, \bibinfo
  {author} {\bibfnamefont {H.-J.}\ \bibnamefont {Eisler}}, \ and\ \bibinfo
  {author} {\bibfnamefont {M.~G.}\ \bibnamefont {Bawendi}},\ }\href@noop {}
  {\bibfield  {journal} {\bibinfo  {journal} {Science},\ }\textbf {\bibinfo
  {volume} {290}},\ \bibinfo {pages} {314} (\bibinfo {year}
  {2000})}\BibitemShut {NoStop}%
\bibitem [{\citenamefont {Michalet}\ \emph {et~al.}(2005)\citenamefont
  {Michalet}, \citenamefont {Pinaud}, \citenamefont {Bentolila}, \citenamefont
  {Tsay}, \citenamefont {Doose}, \citenamefont {Li}, \citenamefont
  {Sundaresan}, \citenamefont {Wu}, \citenamefont {Gambhir},\ and\
  \citenamefont {Weiss}}]{MichaletScience05}%
  \BibitemOpen
  \bibfield  {author} {\bibinfo {author} {\bibfnamefont {X.}~\bibnamefont
  {Michalet}}, \bibinfo {author} {\bibfnamefont {F.~F.}\ \bibnamefont
  {Pinaud}}, \bibinfo {author} {\bibfnamefont {L.~A.}\ \bibnamefont
  {Bentolila}}, \bibinfo {author} {\bibfnamefont {J.~M.}\ \bibnamefont {Tsay}},
  \bibinfo {author} {\bibfnamefont {S.}~\bibnamefont {Doose}}, \bibinfo
  {author} {\bibfnamefont {J.~J.}\ \bibnamefont {Li}}, \bibinfo {author}
  {\bibfnamefont {G.}~\bibnamefont {Sundaresan}}, \bibinfo {author}
  {\bibfnamefont {A.~M.}\ \bibnamefont {Wu}}, \bibinfo {author} {\bibfnamefont
  {S.~S.}\ \bibnamefont {Gambhir}}, \ and\ \bibinfo {author} {\bibfnamefont
  {S.}~\bibnamefont {Weiss}},\ }\href@noop {} {\bibfield  {journal} {\bibinfo
  {journal} {Science},\ }\textbf {\bibinfo {volume} {307}},\ \bibinfo {pages}
  {538} (\bibinfo {year} {2005})}\BibitemShut {NoStop}%
\bibitem [{\citenamefont {Efros}\ \emph {et~al.}(1996)\citenamefont {Efros},
  \citenamefont {Rosen}, \citenamefont {Kuno}, \citenamefont {Nirmal},
  \citenamefont {Norris},\ and\ \citenamefont {Bawendi}}]{EfrosPRB96}%
  \BibitemOpen
  \bibfield  {author} {\bibinfo {author} {\bibfnamefont {A.}~\bibnamefont
  {Efros}}, \bibinfo {author} {\bibfnamefont {M.}~\bibnamefont {Rosen}},
  \bibinfo {author} {\bibfnamefont {M.}~\bibnamefont {Kuno}}, \bibinfo {author}
  {\bibfnamefont {M.}~\bibnamefont {Nirmal}}, \bibinfo {author} {\bibfnamefont
  {D.}~\bibnamefont {Norris}}, \ and\ \bibinfo {author} {\bibfnamefont
  {M.}~\bibnamefont {Bawendi}},\ }\href@noop {} {\bibfield  {journal} {\bibinfo
   {journal} {Phys. Rev. B},\ }\textbf {\bibinfo {volume} {54}},\ \bibinfo
  {pages} {4843} (\bibinfo {year} {1996})}\BibitemShut {NoStop}%
\bibitem [{\citenamefont {Zhao}\ \emph {et~al.}(2007)\citenamefont {Zhao},
  \citenamefont {Graf}, \citenamefont {Jones}, \citenamefont {Franceschetti},
  \citenamefont {Li}, \citenamefont {Wang},\ and\ \citenamefont
  {Kim}}]{ZhaoNL2007}%
  \BibitemOpen
  \bibfield  {author} {\bibinfo {author} {\bibfnamefont {Q.}~\bibnamefont
  {Zhao}}, \bibinfo {author} {\bibfnamefont {P.~A.}\ \bibnamefont {Graf}},
  \bibinfo {author} {\bibfnamefont {W.~B.}\ \bibnamefont {Jones}}, \bibinfo
  {author} {\bibfnamefont {A.}~\bibnamefont {Franceschetti}}, \bibinfo {author}
  {\bibfnamefont {J.}~\bibnamefont {Li}}, \bibinfo {author} {\bibfnamefont
  {L.-W.}\ \bibnamefont {Wang}}, \ and\ \bibinfo {author} {\bibfnamefont
  {K.}~\bibnamefont {Kim}},\ }\href@noop {} {\bibfield  {journal} {\bibinfo
  {journal} {Nano Lett.},\ }\textbf {\bibinfo {volume} {7}},\ \bibinfo {pages}
  {3274} (\bibinfo {year} {2007})}\BibitemShut {NoStop}%
\bibitem [{\citenamefont {Schoenlein}\ \emph {et~al.}(1993)\citenamefont
  {Schoenlein}, \citenamefont {Mittelman}, \citenamefont {Shiang},
  \citenamefont {Alivisatos},\ and\ \citenamefont {Shank}}]{SchoenleinPRL93}%
  \BibitemOpen
  \bibfield  {author} {\bibinfo {author} {\bibfnamefont {R.}~\bibnamefont
  {Schoenlein}}, \bibinfo {author} {\bibfnamefont {D.}~\bibnamefont
  {Mittelman}}, \bibinfo {author} {\bibfnamefont {J.}~\bibnamefont {Shiang}},
  \bibinfo {author} {\bibfnamefont {A.}~\bibnamefont {Alivisatos}}, \ and\
  \bibinfo {author} {\bibfnamefont {C.}~\bibnamefont {Shank}},\ }\href@noop {}
  {\bibfield  {journal} {\bibinfo  {journal} {Phys. Rev. Lett.},\ }\textbf
  {\bibinfo {volume} {70}},\ \bibinfo {pages} {1014} (\bibinfo {year}
  {1993})}\BibitemShut {NoStop}%
\bibitem [{\citenamefont {Mittleman}\ \emph {et~al.}(1994)\citenamefont
  {Mittleman}, \citenamefont {Schoenlein}, \citenamefont {Shiang},
  \citenamefont {Colvin}, \citenamefont {Alivisatos},\ and\ \citenamefont
  {Shank}}]{MittlemanPRB94}%
  \BibitemOpen
  \bibfield  {author} {\bibinfo {author} {\bibfnamefont {D.}~\bibnamefont
  {Mittleman}}, \bibinfo {author} {\bibfnamefont {R.}~\bibnamefont
  {Schoenlein}}, \bibinfo {author} {\bibfnamefont {J.}~\bibnamefont {Shiang}},
  \bibinfo {author} {\bibfnamefont {V.}~\bibnamefont {Colvin}}, \bibinfo
  {author} {\bibfnamefont {A.}~\bibnamefont {Alivisatos}}, \ and\ \bibinfo
  {author} {\bibfnamefont {C.}~\bibnamefont {Shank}},\ }\href@noop {}
  {\bibfield  {journal} {\bibinfo  {journal} {Phys. Rev. B},\ }\textbf
  {\bibinfo {volume} {49}},\ \bibinfo {pages} {14435} (\bibinfo {year}
  {1994})}\BibitemShut {NoStop}%
\bibitem [{\citenamefont {Palinginis}\ \emph {et~al.}(2003)\citenamefont
  {Palinginis}, \citenamefont {Tavenner}, \citenamefont {Lonergan},\ and\
  \citenamefont {Wang}}]{PalinginisPRB03}%
  \BibitemOpen
  \bibfield  {author} {\bibinfo {author} {\bibfnamefont {P.}~\bibnamefont
  {Palinginis}}, \bibinfo {author} {\bibfnamefont {S.}~\bibnamefont
  {Tavenner}}, \bibinfo {author} {\bibfnamefont {M.}~\bibnamefont {Lonergan}},
  \ and\ \bibinfo {author} {\bibfnamefont {H.}~\bibnamefont {Wang}},\
  }\href@noop {} {\bibfield  {journal} {\bibinfo  {journal} {Phys. Rev. B},\
  }\textbf {\bibinfo {volume} {67}},\ \bibinfo {pages} {201307} (\bibinfo
  {year} {2003})}\BibitemShut {NoStop}%
\bibitem [{\citenamefont {Fern\'{e}e}\ \emph {et~al.}(2008)\citenamefont
  {Fern\'{e}e}, \citenamefont {Littleton}, \citenamefont {Cooper},
  \citenamefont {Rubinsztein-Dunlop}, \citenamefont {G\'{o}mez},\ and\
  \citenamefont {Mulvaney}}]{FerneeJPCC08}%
  \BibitemOpen
  \bibfield  {author} {\bibinfo {author} {\bibfnamefont {M.~J.}\ \bibnamefont
  {Fern\'{e}e}}, \bibinfo {author} {\bibfnamefont {B.~N.}\ \bibnamefont
  {Littleton}}, \bibinfo {author} {\bibfnamefont {S.}~\bibnamefont {Cooper}},
  \bibinfo {author} {\bibfnamefont {H.}~\bibnamefont {Rubinsztein-Dunlop}},
  \bibinfo {author} {\bibfnamefont {D.~E.}\ \bibnamefont {G\'{o}mez}}, \ and\
  \bibinfo {author} {\bibfnamefont {P.}~\bibnamefont {Mulvaney}},\ }\href@noop
  {} {\bibfield  {journal} {\bibinfo  {journal} {J. Phys. Chem. C},\ }\textbf
  {\bibinfo {volume} {112}},\ \bibinfo {pages} {1878} (\bibinfo {year}
  {2008})}\BibitemShut {NoStop}%
\bibitem [{\citenamefont {Biadala}\ \emph {et~al.}(2009)\citenamefont
  {Biadala}, \citenamefont {Louyer}, \citenamefont {Tamarat},\ and\
  \citenamefont {Lounis}}]{BiadalaPRL09}%
  \BibitemOpen
  \bibfield  {author} {\bibinfo {author} {\bibfnamefont {L.}~\bibnamefont
  {Biadala}}, \bibinfo {author} {\bibfnamefont {Y.}~\bibnamefont {Louyer}},
  \bibinfo {author} {\bibfnamefont {P.}~\bibnamefont {Tamarat}}, \ and\
  \bibinfo {author} {\bibfnamefont {B.}~\bibnamefont {Lounis}},\ }\Doi
  {10.1103/PhysRevLett.103.037404} {\bibfield  {journal} {\bibinfo  {journal}
  {Phys. Rev. Lett.},\ }\textbf {\bibinfo {volume} {103}},\ \bibinfo {pages}
  {037404} (\bibinfo {year} {2009})}\BibitemShut {NoStop}%
\bibitem [{\citenamefont {Coolen}\ \emph {et~al.}(2008)\citenamefont {Coolen},
  \citenamefont {Brokmann}, \citenamefont {Spinicelli},\ and\ \citenamefont
  {Hermier}}]{CoolenPRL08}%
  \BibitemOpen
  \bibfield  {author} {\bibinfo {author} {\bibfnamefont {L.}~\bibnamefont
  {Coolen}}, \bibinfo {author} {\bibfnamefont {X.}~\bibnamefont {Brokmann}},
  \bibinfo {author} {\bibfnamefont {P.}~\bibnamefont {Spinicelli}}, \ and\
  \bibinfo {author} {\bibfnamefont {J.-P.}\ \bibnamefont {Hermier}},\
  }\href@noop {} {\bibfield  {journal} {\bibinfo  {journal} {Phys. Rev. Lett},\
  }\textbf {\bibinfo {volume} {100}},\ \bibinfo {pages} {027403} (\bibinfo
  {year} {2008})}\BibitemShut {NoStop}%
\bibitem [{\citenamefont {Labeau}\ \emph {et~al.}(2003)\citenamefont {Labeau},
  \citenamefont {Tamarat},\ and\ \citenamefont {Lounis}}]{LabeauPRL03}%
  \BibitemOpen
  \bibfield  {author} {\bibinfo {author} {\bibfnamefont {O.}~\bibnamefont
  {Labeau}}, \bibinfo {author} {\bibfnamefont {P.}~\bibnamefont {Tamarat}}, \
  and\ \bibinfo {author} {\bibfnamefont {B.}~\bibnamefont {Lounis}},\
  }\href@noop {} {\bibfield  {journal} {\bibinfo  {journal} {Phys. Rev.
  Lett.},\ }\textbf {\bibinfo {volume} {90}},\ \bibinfo {pages} {257404}
  (\bibinfo {year} {2003})}\BibitemShut {NoStop}%
\bibitem [{\citenamefont {Borri}\ \emph {et~al.}(2001)\citenamefont {Borri},
  \citenamefont {Langbein}, \citenamefont {Schneider}, \citenamefont {Woggon},
  \citenamefont {Sellin}, \citenamefont {Ouyang},\ and\ \citenamefont
  {Bimberg}}]{BorriPRL01}%
  \BibitemOpen
  \bibfield  {author} {\bibinfo {author} {\bibfnamefont {P.}~\bibnamefont
  {Borri}}, \bibinfo {author} {\bibfnamefont {W.}~\bibnamefont {Langbein}},
  \bibinfo {author} {\bibfnamefont {S.}~\bibnamefont {Schneider}}, \bibinfo
  {author} {\bibfnamefont {U.}~\bibnamefont {Woggon}}, \bibinfo {author}
  {\bibfnamefont {R.~L.}\ \bibnamefont {Sellin}}, \bibinfo {author}
  {\bibfnamefont {D.}~\bibnamefont {Ouyang}}, \ and\ \bibinfo {author}
  {\bibfnamefont {D.}~\bibnamefont {Bimberg}},\ }\href@noop {} {\bibfield
  {journal} {\bibinfo  {journal} {Phys. Rev. Lett.},\ }\textbf {\bibinfo
  {volume} {87}},\ \bibinfo {pages} {157401} (\bibinfo {year}
  {2001})}\BibitemShut {NoStop}%
\bibitem [{\citenamefont {Borri}\ and\ \citenamefont
  {Langbein}(2007)}]{BorriJPCM07}%
  \BibitemOpen
  \bibfield  {author} {\bibinfo {author} {\bibfnamefont {P.}~\bibnamefont
  {Borri}}\ and\ \bibinfo {author} {\bibfnamefont {W.}~\bibnamefont
  {Langbein}},\ }\href@noop {} {\bibfield  {journal} {\bibinfo  {journal} {J.
  Phys.: Condens. Matter.},\ }\textbf {\bibinfo {volume} {19}},\ \bibinfo
  {pages} {295201} (\bibinfo {year} {2007})}\BibitemShut {NoStop}%
\bibitem [{\citenamefont {Shah}(1996)}]{ShahBook96}%
  \BibitemOpen
  \bibfield  {author} {\bibinfo {author} {\bibfnamefont {J.}~\bibnamefont
  {Shah}},\ }\enquote {\bibinfo {title} {Ultrafast spectroscopy of
  semiconductors and semiconductor nanostructures},}\ \ (\bibinfo  {publisher}
  {Springer},\ \bibinfo {address} {Berlin},\ \bibinfo {year} {1996})\
  Chap.~\bibinfo {chapter} {2}\BibitemShut {NoStop}%
\bibitem [{\citenamefont {Masia}\ \emph {et~al.}(2011)\citenamefont {Masia},
  \citenamefont {Langbein}, \citenamefont {Moreels}, \citenamefont {Hens},\
  and\ \citenamefont {Borri}}]{MasiaPRB11}%
  \BibitemOpen
  \bibfield  {author} {\bibinfo {author} {\bibfnamefont {F.}~\bibnamefont
  {Masia}}, \bibinfo {author} {\bibfnamefont {W.}~\bibnamefont {Langbein}},
  \bibinfo {author} {\bibfnamefont {I.}~\bibnamefont {Moreels}}, \bibinfo
  {author} {\bibfnamefont {Z.}~\bibnamefont {Hens}}, \ and\ \bibinfo {author}
  {\bibfnamefont {P.}~\bibnamefont {Borri}},\ }\href@noop {} {\bibfield
  {journal} {\bibinfo  {journal} {Phys. Rev. B.},\ }\textbf {\bibinfo {volume}
  {83}},\ \bibinfo {pages} {201309(R)} (\bibinfo {year} {2011})}\BibitemShut
  {NoStop}%
\bibitem [{\citenamefont {Borri}\ \emph {et~al.}(2005)\citenamefont {Borri},
  \citenamefont {Langbein}, \citenamefont {Woggon}, \citenamefont {Stavarache},
  \citenamefont {Reuter},\ and\ \citenamefont {Wieck}}]{BorriPRB05}%
  \BibitemOpen
  \bibfield  {author} {\bibinfo {author} {\bibfnamefont {P.}~\bibnamefont
  {Borri}}, \bibinfo {author} {\bibfnamefont {W.}~\bibnamefont {Langbein}},
  \bibinfo {author} {\bibfnamefont {U.}~\bibnamefont {Woggon}}, \bibinfo
  {author} {\bibfnamefont {V.}~\bibnamefont {Stavarache}}, \bibinfo {author}
  {\bibfnamefont {D.}~\bibnamefont {Reuter}}, \ and\ \bibinfo {author}
  {\bibfnamefont {A.}~\bibnamefont {Wieck}},\ }\href@noop {} {\bibfield
  {journal} {\bibinfo  {journal} {Phys. Rev. B},\ }\textbf {\bibinfo {volume}
  {71}},\ \bibinfo {pages} {115328} (\bibinfo {year} {2005})}\BibitemShut
  {NoStop}%
\bibitem [{Not(){\natexlab{a}}}]{NoteCdSe}%
  \BibitemOpen
  \href@noop {} {} {\natexlab{a}}\ \bibinfo {note} {Near the cross-over, it
  is also possible that the $0^{\rm L}$ dark state is the lowest dark, and just
  above the $|F|=1$ bright state there is a dark $|F|=2$ or a bright $0^{\rm
  U}$ state\,\cite{EfrosPRB96,ZhaoNL2007}, the latter possibly contributing to
  the sub-ps dynamics previously discussed.}\BibitemShut {Stop}%
\bibitem [{Not(){\natexlab{b}}}]{NoteCdSe2}%
  \BibitemOpen
  \href@noop {} {} {\natexlab{b}}\ \bibinfo {note} {A good fit as shown in
  Fig.\,\ref{Fig2}b is also achieved considering the lowest level $0^{\rm L}$
  dark and the upper level $0^{\rm U}$ bright, with ${\Delta}E_{\rm D}=1.7 \pm
  0.2$\,meV and $\gamma_{\rm B} = 1.32 \pm 0.03 \times 10^{-1}
  \,\textrm{ns}^{-1}$}\BibitemShut {NoStop}%
\bibitem [{\citenamefont {Langbein}\ \emph {et~al.}(1997)\citenamefont
  {Langbein}, \citenamefont {Hvam}, \citenamefont {Umlauff}, \citenamefont
  {Kalt}, \citenamefont {Jobst},\ and\ \citenamefont {Hommel}}]{LangbeinPRB97}%
  \BibitemOpen
  \bibfield  {author} {\bibinfo {author} {\bibfnamefont {W.}~\bibnamefont
  {Langbein}}, \bibinfo {author} {\bibfnamefont {J.~M.}\ \bibnamefont {Hvam}},
  \bibinfo {author} {\bibfnamefont {M.}~\bibnamefont {Umlauff}}, \bibinfo
  {author} {\bibfnamefont {H.}~\bibnamefont {Kalt}}, \bibinfo {author}
  {\bibfnamefont {B.}~\bibnamefont {Jobst}}, \ and\ \bibinfo {author}
  {\bibfnamefont {D.}~\bibnamefont {Hommel}},\ }\href@noop {} {\bibfield
  {journal} {\bibinfo  {journal} {Phys. Rev. B},\ }\textbf {\bibinfo {volume}
  {55}},\ \bibinfo {pages} {R7383} (\bibinfo {year} {1997})}\BibitemShut
  {NoStop}%
\bibitem [{\citenamefont {Langbein}\ \emph {et~al.}(2004)\citenamefont
  {Langbein}, \citenamefont {Borri}, \citenamefont {Woggon}, \citenamefont
  {Stavarache}, \citenamefont {Reuter},\ and\ \citenamefont
  {Wieck}}]{LangbeinPRB04a}%
  \BibitemOpen
  \bibfield  {author} {\bibinfo {author} {\bibfnamefont {W.}~\bibnamefont
  {Langbein}}, \bibinfo {author} {\bibfnamefont {P.}~\bibnamefont {Borri}},
  \bibinfo {author} {\bibfnamefont {U.}~\bibnamefont {Woggon}}, \bibinfo
  {author} {\bibfnamefont {V.}~\bibnamefont {Stavarache}}, \bibinfo {author}
  {\bibfnamefont {D.}~\bibnamefont {Reuter}}, \ and\ \bibinfo {author}
  {\bibfnamefont {A.~D.}\ \bibnamefont {Wieck}},\ }\href@noop {} {\bibfield
  {journal} {\bibinfo  {journal} {Phys. Rev. B},\ }\textbf {\bibinfo {volume}
  {70}},\ \bibinfo {pages} {033301} (\bibinfo {year} {2004})}\BibitemShut
  {NoStop}%
\end{thebibliography}

%

\end{document}